# Dynamic sea level changes following changes in the thermohaline circulation

Anders Levermann[1], Alexa Griesel[1], Matthias Hofmann[1], Marisa Montoya[2] & Stefan Rahmstorf[1]

[1]*Potsdam Institute for Climate Impact Research, 14473 Potsdam, Germany*

[2]*Facultad de Ciencias Fisicas, Universidad Complutense de Madrid, Madrid, Spain*

**Abstract** Using the coupled climate model CLIMBER-3$\alpha$, we investigate changes in sea surface elevation due to a weakening of the thermohaline circulation (THC). In addition to a global sea level rise due to a warming of the deep sea, this leads to a regional dynamic sea level change which follows quasi-instantaneously any change in the ocean circulation. We show that the magnitude of this dynamic effect can locally reach up to ~1m, depending on the initial THC strength. In some regions the rate of change can be up to 20-25 mm/yr. The emerging patterns are discussed with respect to the oceanic circulation changes. Most prominent is a south-north gradient reflecting the changes in geostrophic surface currents. Our results suggest that an analysis of observed sea level change patterns could be useful for monitoring the THC strength.

## 1 Introduction

Sea level in the northern Atlantic is significantly lower compared to the northern Pacific (see Figure 1a) as a result of deep water formation in the high latitudes of the North Atlantic; this leads, e.g., to a flow through Bering Strait from the Pacific to the Arctic Ocean (WIJFFELS et al., 1992). Hence, any major change in North Atlantic deep water formation can be expected to cause important sea level changes around the northern

Atlantic. This is a consequence of THC changes that is often overlooked (VELLINGA and WOOD, 2002; SCHWARTZ and RANDALL, 2003).

Paleoclimatic data indicate that strong variations of the THC have occurred in the past (see references in (CLARK et al., 2002; RAHMSTORF, 2002)). Model simulations suggest that global warming accompanied by an increased freshwater input into the North Atlantic could result in major changes in THC strength (MANABE and STOUFFER, 1993; STOCKER and SCHMITTNER, 1997; RAHMSTORF, 1999; RAHMSTORF and GANOPOLSKI, 1999; WOOD et al., 1999; 2001; SCHAEFFER et al., 2002). Recent observations show an increase by 7% in river discharge to the Arctic Ocean (PETERSON et al., 2002) between 1936 and 1999 and a significant freshening of the Atlantic Ocean during the last 40 years (DICKSON et al., 2002; CURRY et al., 2003). Oceanic section data from the northern Atlantic suggest (LHERMINIER and TEAM, 2004) a weakening of the overturning by 40% between 1997 and 2002. Exploring possible consequences of and indicators for thermohaline circulation changes is hence of great importance. Here, we study the sea level effects of THC changes in a series of simulations with a coupled climate model that contains an ocean circulation model with explicit calculation of sea surface elevation.

It is useful to distinguish three types of sea level changes (in addition to those resulting from tectonic processes, which we will not consider here):

(i) global sea level change due to changes in the amount of water in the oceans, e.g. caused by the melting of continental ice;

(ii) global and regional sea level change due to changes in specific volume of sea water, caused by diabatic heating;

(iii) regional sea level changes (with near-zero global mean) associated with changing ocean currents and a redistribution of mass in the ocean.

A change in thermohaline ocean circulation will directly affect sea level through mechanisms (ii) and (iii). Model simulations suggest that following a shutdown of North Atlantic deep water (NADW) formation, the deep ocean would warm up on a slow diffusive time scale (millennia) as the supply of cold waters of polar origin to the deep ocean is reduced; this leads to a slow global sea level rise (BRYAN, 1996; GREGORY and LOWE, 2000; KNUTTI and STOCKER, 2000; GREGORY et al., 2001; 2001; SEIDOV et al., 2001) of the order of 1m. If NADW formation shuts down in scenarios for future global warming, this sea level rise comes in addition to the rise resulting from the warming of the surface climate.

A change in thermohaline circulation is also associated with a change in ocean currents and a redistribution in mass (i.e., adiabatic adjustments in the density structure of the ocean). This hardly affects the global mean sea level (since the volume of sea water is not changed, except for small differential compressibility effects in the equation of state), but it can lead to rapid regional sea level changes. The time scale of these is set by the adjustment of the internal density structure of the ocean due to internal wave processes (decades). The adjustment of the sea surface itself to changes in circulation and density structure occurs almost instantaneously (time-scale of days) through surface gravity waves.

The sea level gradients are in equilibrium with the surface currents and winds as given by the momentum balance in the upper oceanic layer

$$\rho \cdot f \cdot v = g \cdot \partial_x(\rho\eta) - \partial_z \tau_x \ , \qquad (1\ a)$$

$$\rho \cdot f \cdot u = -g \cdot \partial_y(\rho\eta) + \partial_z \tau_y, \qquad (1\ b)$$

where $\eta$ is the sea surface elevation, and $\rho$, $f$ and $g$ are the density, Coriolis parameter and gravitational constant. The stress field is ($\tau_x, \tau_y$) with the wind stress field ($\tau_{ox}, \tau_{oy}$) as its upper boundary values and $u$ and $v$ the horizontal velocities in the ocean near the surface. In order to isolate and study the sea level effects of THC changes, we keep the wind stress constant throughout our experiments (prescribed from NCEP/NCAR reanalysis data (KALNAY et al., 1996)) despite otherwise using a fully coupled ocean-atmosphere model.

## 2 Model and experimental set-up

The simulations discussed here were carried out using the coupled climate model CLIMBER-3α, which contains an improved version of the oceanic general circulation model MOM-3 and thereby a non-linear explicit free surface representation. The model includes interactive atmosphere (PETOUKHOV et al., 2000) and sea ice modules (FICHEFET and MAQUEDA, 1997). The wind stress is prescribed from the NCEP-NCAR reanalysis (KALNAY et al., 1996), so that the modelled sea level changes are entirely due to ocean dynamics as opposed to wind stress changes. The resolution is 3.75°x3.75° horizontally with 24 vertical levels. For a full description of the model see (ref. (MONTOYA et al., 2004 (in preparation))).

Figure 1 shows a comparison of altimeter data from the TOPEX-POSEIDON satellite (NOAA, 1992-1995) of the deviation of the sea surface elevation from the global mean, with the free surface simulation of the present-day climate with CLIMBER-3α. The model reproduces the main features of the sea level patterns. The strongest differences can be found in the Indian Ocean and along the South American East coast where too fresh surface water leads to an unrealistically high $\eta$ in the model. These problems are not severe, considering that this is a thermodynamic equilibrium of a coarse-resolution model without flux adjustments of heat or freshwater.

The corresponding meridional streamfunction (Figure 2a) for the same present-day equilibrium shows 12 Sv (1 Sv = $10^6$ m$^3$ s$^{-1}$) overturning of NADW, a value at the lower end of observational estimates (GANACHAUD and WUNSCH, 2000). Figure 3 shows the mixed layer depth of the simulation. The deep water formation sites are visible as areas of deep mixed layers. NADW production takes place in GIN Sea and Irminger Sea, while Antarctic Bottom Water is formed near the Antarctic coast in Weddell Sea and Ross Sea.

The deep convection sites can be seen as associated sea level minima in Figure 1b. The sea level difference between northern Atlantic and Pacific is 57 cm in the northernmost available strip of the altimeter data (i.e., averaged across each basin between 55ºN and 65ºN). In the model, this difference is underestimated as 34 cm, which is most likely due to the weak overturning; we will see later on that in the model this sea level difference is roughly proportional to the Atlantic overturning rate. Hence, our results for the sea level changes resulting from a THC shutdown should be considered as a conservative lower limit.

Starting from this present-day equilibrium state, we apply a constant amount of negative salt flux to the two main convection sites in the North Atlantic (52°N to 80°N by 48°W to 15°E, excluding regions of shallow water along the northern European coast at 52°N to 64°N by 12°W to 8°E and along the Greenland coast at 63°N to 72°N by 31°W to 22°W), in order to cause a weakening of the overturning. A salt (rather than freshwater) flux was applied in order to avoid a sea level change due to adding water, which in the free surface formulation of our model would directly raise sea level. The THC strength decreases with increasing negative salt flux and collapses if the chosen salt flux is larger than 0.3 Sv freshwater equivalent. We used salt fluxes equivalent to 0.05, 0.075, 0.1, 0.15, 0.2, 0.25, 0.3, 0.35 and 0.4 Sv of freshwater flux which were applied until quasi-equilibrium was reached after 800 years.

The control state simulation presented here was chosen to give the most realistic oceanic circulation as well as tracer distribution. Our results are robust with respect to changes in model parameters such as vertical diffusivity. The same experiments were carried out with the MOM-3 ocean model with a slightly coarser horizontal resolution of 4°x4° coupled to an energy-moisture balance model and the same sea ice model. With different oceanic mixing parameterization, this model is bistable, showing a stable THC off-state in addition to the on-state for present-day climate (as in ref. (MANABE and STOUFFER, 1988)). In this case the salt flux was removed after a brief pulse equivalent to 1 Sv for 50 years. During the successive integration over 800 years without salt flux forcing, the sea level patterns did not change significantly. The patterns show the same structure and magnitude as the simulations shown here.

Figure 2b shows the streamfunction after the application of 0.35 Sv freshwater equivalent salt flux for 800 years. In this model state, the sea level difference between northern Atlantic and Pacific is –7 cm, i.e., has almost vanished.

## 3  Time series of dynamic sea level changes associated with THC break-down

Figure 4a shows the maximum of the Atlantic overturning circulation as a function of time for a salt flux equivalent of 0.35 Sv (black line, right axis). Starting from a strength of 12 Sv, the salt flux was applied from time t=0 onwards, causing a rapid decline of the overturning strength and its practical disappearance after 100 years. Also plotted in Figure 4a are the corresponding changes in dynamic sea level for different locations in the North Atlantic and the Southern Ocean (left axis). For the annual average values plotted in the figure, the response of the dynamic sea level is synchronous in the North Atlantic and in the Southern Ocean due to the fast travelling speed of the gravity waves that transport the sea level signal. Sea level rise is strongest on the coasts of Europe and North America. The maximum value of 50 cm depends on the initial strength of the THC; a stronger sea level signal would result for a stronger initial overturning strength. Figure 4a shows that a rapid initial sea level rise occurs within decades after the onset of the THC collapse. The rise in the North Atlantic is accompanied by a rapid decline of sea level in the Southern Ocean; the global integral of the sea level curves shown is zero, since the volume of water in the global ocean is conserved during the experiments and changes in mean specific volume (mechanism (ii) discussed in the introduction) are *not* included in the model. The latter global mean change in sea level is plotted separately in Figure 4a (labelled "thermal expansion"); the full sea level change is the sum of both effects (GREATBATCH, 1994).

Figure 4b gives the *rate* of change of sea level difference for the region where the Gulf Stream travels along the North American coast for different values of the salt flux applied to the northern convection sites. The rate of change varies for freshwater equivalents up to 0.3 Sv, for which the THC has weakened but not completely collapsed. For larger forcing the THC collapses and the rate of change saturates into a single curve with a maximum around 25 mm y$^{-1}$.

## 4  Patterns of dynamic sea level change

Figure 5 shows the two-dimensional pattern of the dynamic sea level change (again, the global mean of this pattern is zero). The pattern shown is the equilibrium reached after the THC shutdown after 800 years integration. The sea level changes reflect the changes in the geostrophic surface currents as described in equations (1a,b). The ageostrophic velocities are negligible outside a strip of 3.75° (one model grid cell) around the equator; away from the equator the ratio of ageostrophic to geostrophic velocities is less then 1%. The wind stress was taken from the NCEP-NCAR reanalysis for all simulations, i.e. sea level differences discussed in this article are entirely due to oceanic circulation changes.

On top of the sea level differences, Figure 5 shows the difference in surface currents between the THC off-state and the THC on-state. The dominant feature in the pattern is a south-north sea level gradient, reflecting the zonal components of the upper branch of the THC. In the control run, the current crosses the Atlantic from East to West between 30°S and 40°S associated with a northward drop in sea surface elevation. The elimination of this current during a THC shutdown yields a northward sea level rise as seen in Figure 5. This effect is even clearer in the North Atlantic (around 40°N) where the Gulf Stream crosses the Atlantic basin in order to enter the GIN Sea as the North Atlantic Current. The stoppage of the North Atlantic Current results in the strong northward gradient of sea level difference around 40°N and the strong onshore gradient along the North American coast seen in the figure.

The Northern convection sites appear as distinct locations of sea level rise, reflecting the elimination of deep down-welling there. A strengthening of the Antarctic Circumpolar Current of about 8 Sv is seen as a sea level drop in the Southern Ocean around 60°S. Note that the formation of Antarctic Bottom Water in the present day equilibrium simulation takes place mainly through convection in the Weddell Sea and does not directly relate to the decline of sea level in the Southern Ocean.

The patterns seen in Figure 5 result from the difference in sea level between two equilibrium runs; one with and the other without additional salt flux to the northern convection sites. To confirm that the pattern obtained is robust and due to the THC change (not a direct consequence of the salt forcing), we performed a number of sensitivity experiments with varying parameter choices and with other models. When coupling the same ocean module to an energy-moisture-balance model, the application of a freshwater pulse to the convection sites shuts the THC down permanently; i.e., it remains off after removal of the forcing. The sea level changes found in this off-state were of the same order of magnitude as the ones shown in Figure 5. The effect of the additional salt-flux to the northern convection sites on the sea level patterns and magnitudes in the presented simulations is negligible. The qualitative patterns and magnitudes of sea level difference shown were tested to be robust to changes in the parameter set-ups of the model. The simulations shown here use the parameter choice that best reproduces the present ocean circulation.

## 5  Sea level difference and THC strength

Figure 6 shows the sea level change for the region where the Gulf Stream passes the North Atlantic coast and for the European coast as a function of the maximum overturning. The data plotted were obtained by equilibrium experiments with different amounts of salt flux applied to the North Atlantic convection sites. We get a linear dependence of the sea level difference in the North American Gulf Stream region and the strength of the THC with a slope of $\gamma = 5$ cm $Sv^{-1}$. This slope can be understood through the geostrophic balance in equations (2). The strength of the THC, $V$, can be estimated by integrating the northward velocity over a vertical box perpendicular to the North American coast. The integration yields a relation of the sea level drop and the overturning strength $\Delta\eta = \gamma \cdot V$. A rough estimate assuming constant flow for the entire profile of the Gulf Stream which reaches down to a depth of about H=500 m results in a proportionality factor of $\gamma = f / g \cdot H \sim 10^{-4} s^{-1} / 10 ms^{-2} \cdot 500m = 2$ cm $Sv^{-1}$. Refining the estimate by assuming a quadratic velocity profile in the horizontal and in the vertical direction yields a correction of a factor of 2/3 for both directions. Thus the geostrophic slope of the red curve in Figure 6 can be estimated to be about $\gamma = 3/2 \cdot 3/2 \cdot 2$ cm $Sv^{-1} = 4.5$ cm $Sv^{-1}$, in agreement with the slope found in the model. The scaling of sea level difference with THC strength suggest that stronger initial THC strength than the rather low value of 12 Sv of our simulations would lead to a proportionally stronger sea level response if the THC shuts down. Observations (GANACHAUD and WUNSCH, 2000; TALLEY et al., 2003) suggest higher present values for the maximum of the overturning in the Atlantic of about 16-18 Sv. Using the slope of 5 cm Sv$^{-1}$, this would lead to a sea level rise along the North American coast of about 80-90 cm. In order to check the effect of a strengthened THC, we carried out simulations with increased vertical diffusivity. A model run with a vertical diffusivity of 0.4 cm$^2$ s$^{-1}$ (as opposed to 0.1 cm$^2$ s$^{-1}$ for the presented simulations) resulted in a stronger THC of 15.5 Sv and a sea level difference of 82 cm along the North American coast.

## 6  Implications for future climate change and THC monitoring

We have shown that a major change in thermohaline ocean circulation would be associated with large and potentially rapid regional sea level changes due to dynamic adjustment of the ocean surface. A complete cessation of NADW formation, for example, could raise sea level along the North American Atlantic coast by between half a meter and a meter, on a time scale that instantaneously follows the change in ocean circulation. In addition, a cessation of NADW formation would lead to a slow warming of the deep ocean, causing additional sea level rise by several decimetres, albeit on a much slower time scale. While the latter effect was discussed in the 2001 IPCC report (IPCC, 2001), the former, potentially more hazardous one has thus far been largely overlooked.

How anthropogenic climate change will affect the ocean circulation is difficult to predict, both due to uncertainties in the freshwater budget (e.g., the amount of Greenland melt-water runoff) and due to a strong model-dependence in the stability of the ocean circulation which is not yet understood. Studies with coupled atmosphere-ocean general circulation models (MANABE and STOUFFER, 1993; STOCKER and SCHMITTNER, 1997; MANABE and STOUFFER, 1999; RAHMSTORF, 1999; RAHMSTORF and GANOPOLSKI, 1999; WOOD et al., 1999; 2001; STOUFFER and MANABE, 2003) as well as paleodata (RAHMSTORF, 2002) and recent observations (DICKSON et al., 2002;

PETERSON et al., 2002; CURRY et al., 2003) suggest that a weakening or even a temporary or permanent cessation of NADW formation is within the range of possibilities.

Hence, observational monitoring of trends in the THC is important, and efforts to deploy moored arrays are in place or underway (HIRSCHI et al., 2003). The specific regional pattern of sea level changes found here suggests that sea level data could be useful for monitoring changes in THC. The amplitude of 5 cm sea level rise inshore of the Gulf Stream for each Sv of weakening should be detectable. First encouraging results in this direction were presented recently (HÄKKINEN and RHINES, 2004) showing a weakening of the subpolar gyre. Recent altimeter data show a significant sea level rise along the North American coast that can be explained neither by freshwater input nor by ocean warming (CAZENAVE et al., 2004 (submitted)). Whether these regional deviations from the global mean sea level trend are due to changes in the THC strength needs to be clarified by further studies.

**Acknowledgments**  The authors are grateful to J. Mignot for fruitful discussions. A.L. was funded by the Comer foundation. A.G. and M. H. were funded through the James S. McDonnell Foundation Centennial Fellowship. M.M. was funded by the Spanish Ministry for Science and Education through the Ramon y Cajal programme.

**Correspondence**  Anders Levermann, Potsdam Institute for Clmate Impact Research, Telegraphenberg A26, 14473 Potsdam, Germany, e-mail: anders.levermann@pik-potsdam.de.

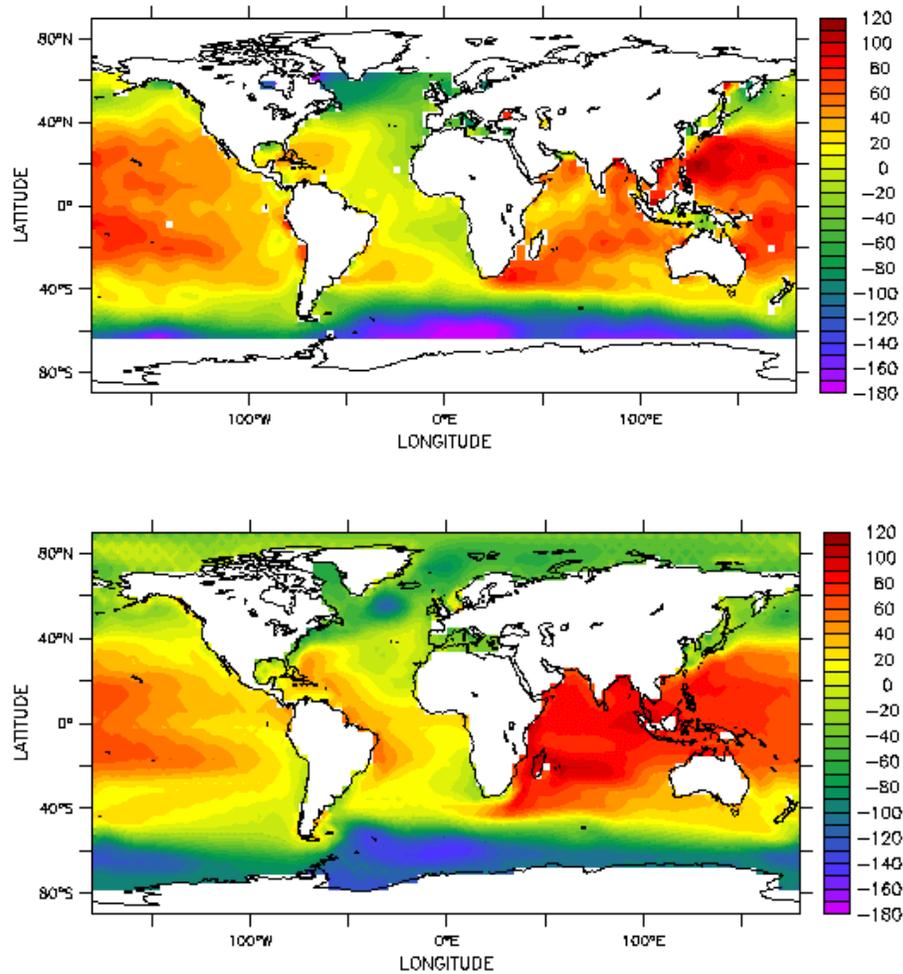

**Fig. 1:** Deviations of sea surface elevation from the global mean (in cm), **(a)** derived from altimeter data from the TOPEX/POSEIDON satellite (NOAA, 1992-1995) in comparison with **(b)** a simulation of the free surface in the coupled climate model CLIMBER-3α (MONTOYA et al., 2004 (in preparation)). Note that areas of deep convection in the northern Atlantic are associated with minima in sea level; in the altimeter data this is the Labrador Sea, while in the model it is the Irminger Sea and the Greenland-Norwegian Sea.

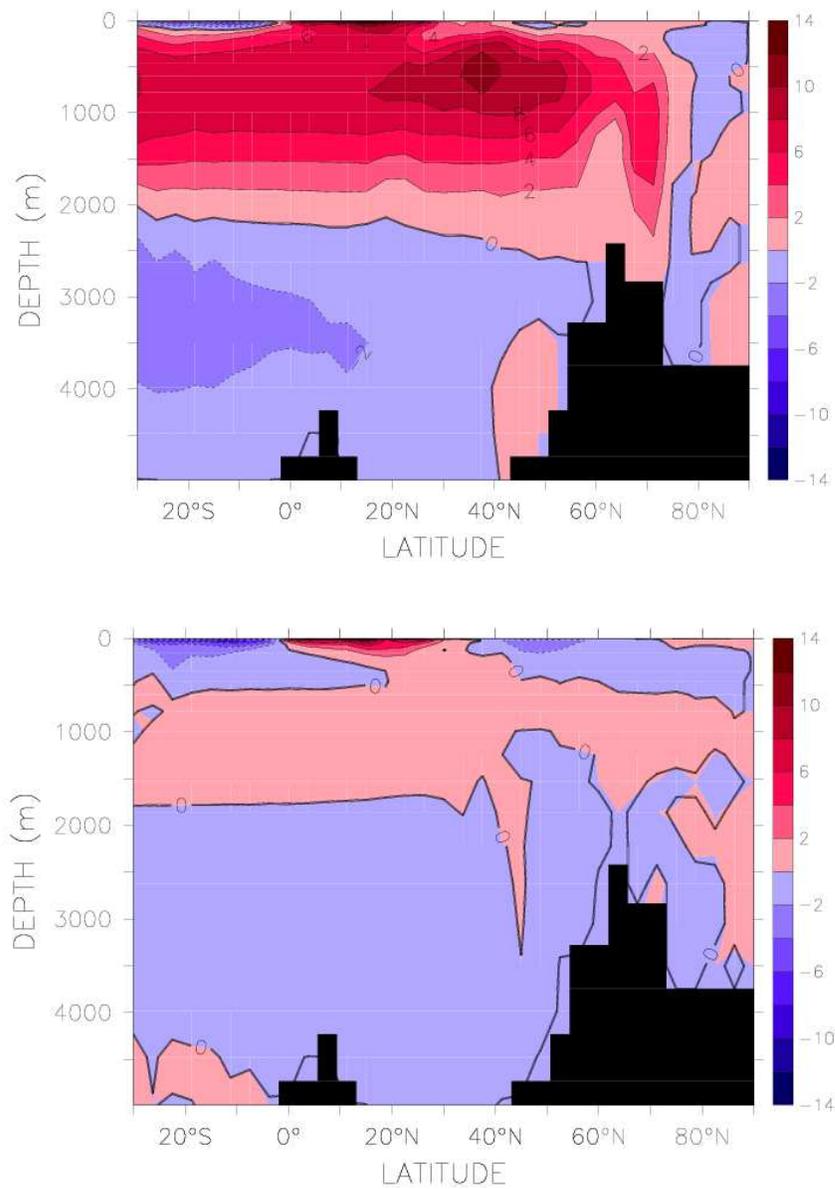

**Fig. 2:** Streamfunction of the meridional overturning circulation in the Atlantic Ocean. (Contour line differences are 2 Sv) **(a)** overturning cell with maximal volume transport of 12 Sv for the present day equilibrium simulation. **(b)** equilibrium with an equivalent freshwater flux of 0.35 Sv, for which no overturning circulation can be sustained.

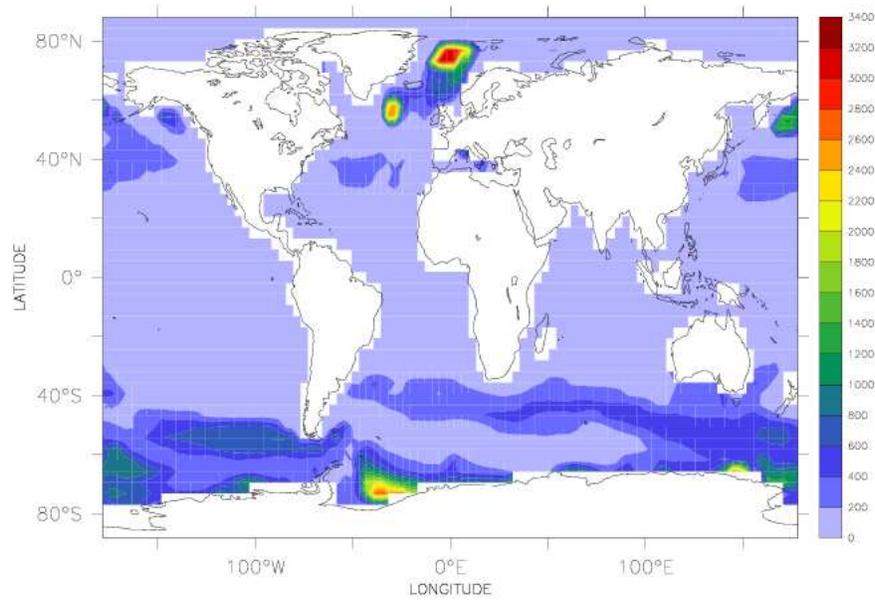

**Fig. 3:** Mixed layer depth (in m) for the control run. Deep water formation sites are the GIN Sea and Irminger Sea for the North Atlantic Deep Water and Weddell Sea and Ross Sea for Antarctic Bottom Water.

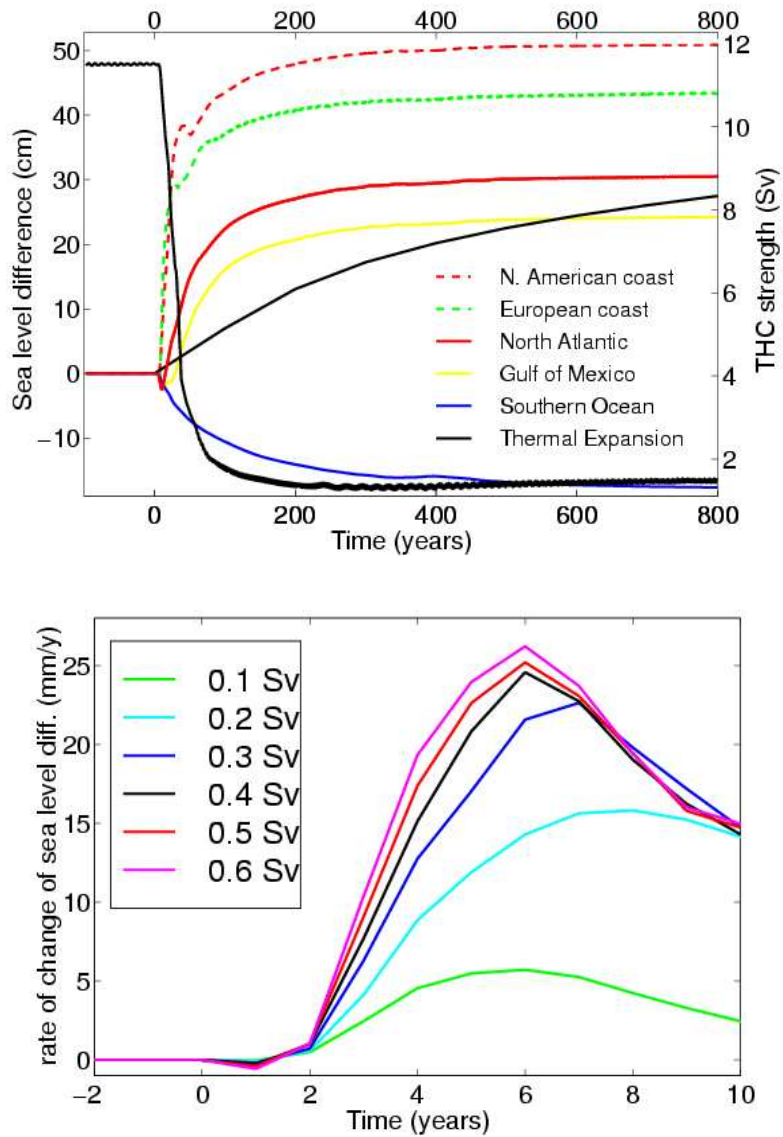

**Fig. 4: (a)** Time evolution (10 year running average) of the THC strength (black, right axis) and corresponding dynamic sea level changes at different locations in the North Atlantic and the Southern Ocean (left axis). Note that all regions are outside the domain of salt flux application. Sea level change due to thermal expansion is also shown. **(b)** Rate of change in sea level along the North

American coast for 10 year running averages and for different values of salt flux forcing. Rates of change increase for stronger weakening of the THC, but saturate after the THC collapse occurs for more than 0.3 Sv of freshwater equivalent forcing.

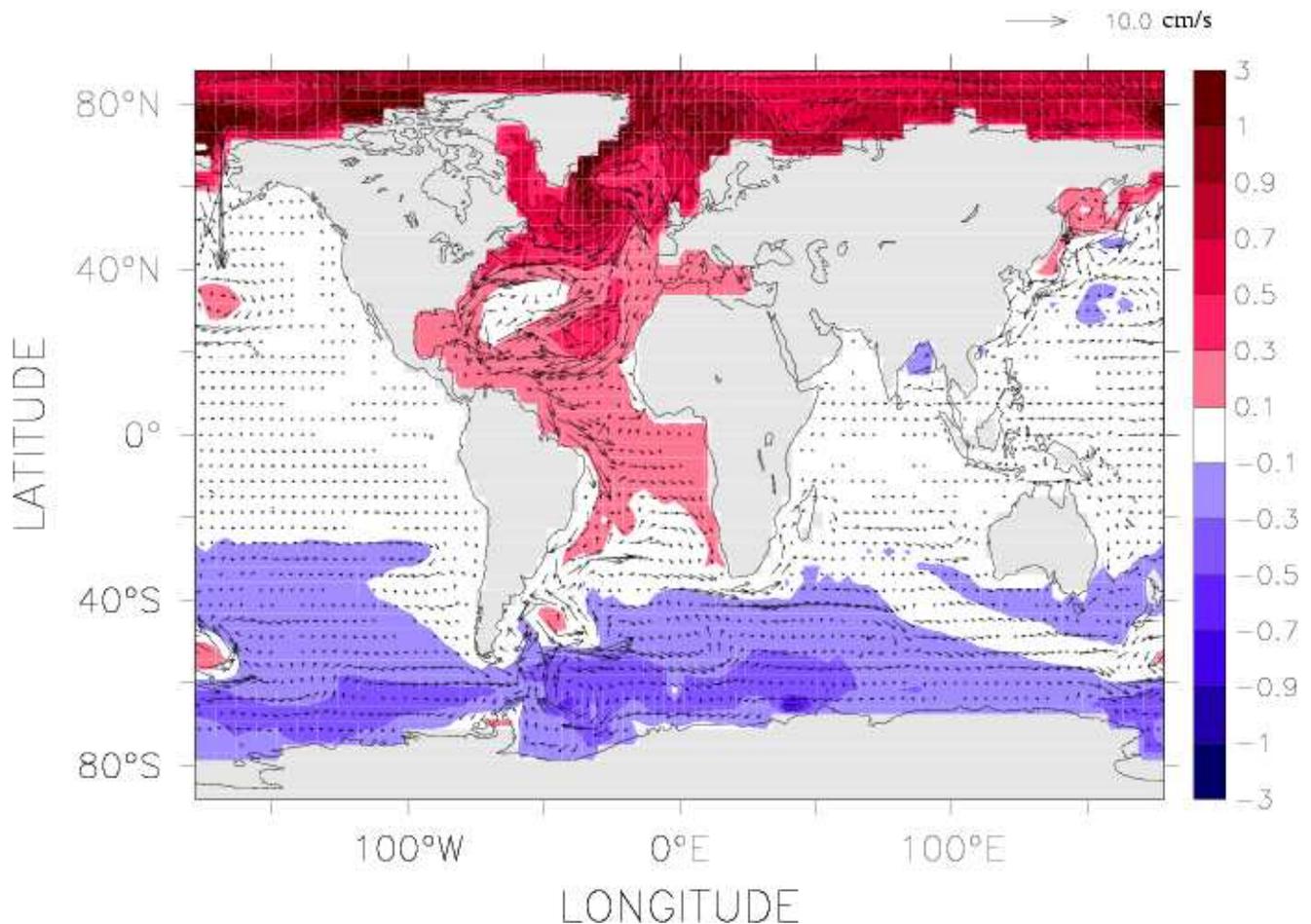

**Fig. 5:** Dynamic sea level changes (in m) after the collapse of the THC (contours), and difference of surface current velocities (arrow in the upper right corner corresponds to 10 cm s$^{-1}$).

17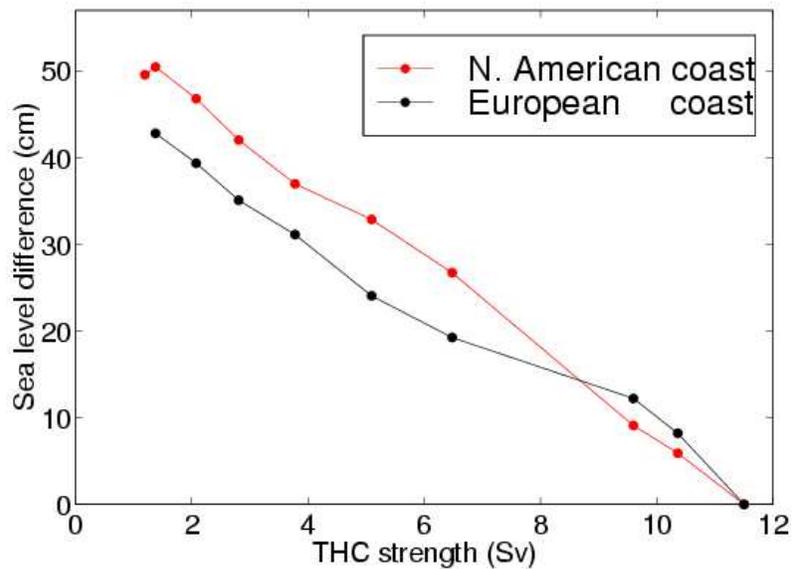

**Fig. 6:** Dynamic sea level changes (in cm) as a function of THC strength (in Sv) for different locations in the North Atlantic.

BRYAN, K. (1996): The steric component of sea level rise associated with enhanced greenhouse warming: a model study. Climate Dynamics **12**: 545.

CAZENAVE, A., A. LOMBARD, K. DOMINH, C. CABANES and R. S. NEREM (2004 (submitted)): Sea level variability, tide gauges, altimetry and thermal expansion. Geophysical Research Letters.

CLARK, P. U., N. G. PISIAS, T. F. STOCKER and A. J. WEAVER (2002): The role of the thermohaline circulation in abrupt climate change. Nature **415**: 863.

CURRY, R., B. DICKSON and I. YASHAYAEV (2003): A change in the freshwater balance of the Atlantic Ocean over the past four decades. Nature **426**: 826.